\documentclass[aps,prb,superscriptaddress,twocolumn,showkeys,amsmath,amssymb,floatfix,longbibliography]{revtex4-2}

\usepackage{comment}
\usepackage[normalem]{ulem}
\usepackage[utf8]{inputenc}
\usepackage{graphicx}
\usepackage{dcolumn}
\usepackage{bm}
\usepackage{hyperref}
\usepackage{amsmath}
\usepackage{mathtools}
\usepackage[dvipsnames]{xcolor}
\usepackage{amssymb}
\usepackage{outline}
\definecolor{darkblue}{RGB}{0,0,139} 

\usepackage{ulem}

\usepackage{xcolor}

\begin{document}

\title{Temperature and spatial variation of topological defects density during the crystallization of  vortex nanocrystals}

\author{M. I. Dolz}
\affiliation{Facultad de Ciencias Físico, Matemáticas y Naturales, Universidad Nacional de San Luis. Instituto de Física Aplicada, CONICET. San Luis, Argentina}

\author{A. B. Kolton}
\affiliation{Condensed Matter Theory group, Centro At\'{o}mico Bariloche, CNEA,  Argentina}
\affiliation{Instituto Balseiro,
CNEA and Universidad Nacional de Cuyo, 
Bariloche, Argentina.}

\author{Y. Fasano}
\affiliation{Instituto Balseiro,
CNEA and Universidad Nacional de Cuyo, 
Bariloche, Argentina.}
\affiliation{Low Temperatures Lab, Centro At\'{o}mico Bariloche, CNEA, Argentina.}
\affiliation{Instituto de Nanociencia y Nanotecnología, CONICET-CNEA, Nodo Bariloche,  Argentina.}

\date{\today}

\begin{abstract}

Understanding the role of confinement while crystallizing  nanocrystals is very relevant for predicting their structure and physical properties. With this aim we perform Langevin dynamics simulations of nanocrystals of the model system of few hundred vortices nucleated in micron-sized superconductors. We study the crystallization dynamics and the low-temperature structural properties of vortex nanocrystals nucleated in field-cooling conditions when changing vortex density or elasticity of the system and physical size of the samples. The low-temperature snapshots obtained in simulations present a healing effect at the edges that is in quantitative agreement with experimental data in  Bi$_2$Sr$_2$CaCu$_2$O$_{8+\delta}$ micron-sized samples.  We show that the low-temperature radial distribution of topological defects is a stationary profile frozen at a temperature below the melting line tuned by intrinsic properties of the vortex structure and on the confinement effect. These findings on the dynamics and spatial profile of topological defects  can be applied to describe the physical properties of confined soft condensed matter nanocrystals in general.
\end{abstract}


\maketitle

\section*{Introduction}

The spatial distribution and dynamics of topological defects in crystallization processes following a particular thermal and deformation history constrains the physical properties of a material and their potential applications.~\cite{Sethna2017}
While systems at equilibrium forget their thermal and deformation history, crystallizing soft condensed matter systems in a substrate with disorder   generates plastic deformations and since the material is not in global equilibrium its properties depend on history.~\cite{Sethna2017,Anderson2002} 
Thermal history is tuned by cooling or warming at different temperature rates; deformation history can be controlled by introducing stress either with an external field or by the edges of the material. For nanoscale condensed matter systems composed by only few thousands of interacting objects, 
the deformation effect introduced by the confinement produced by the edges of the sample becomes relevant  and generates additional topological defects in the nanocrystal.~\cite{Taloni2018}

In general, the density and nature of topological defects 
 can control the occurrence of phase transitions,~\cite{Nelson2002} and the dynamics of interfaces composed of topological defects control the nucleation and size of grains in polycrystalline solids~\cite{Hillert1965,Moretti2005,Salvalaglio2018,Salvalaglio2019,Qiu2024}, which in turn affects the mechanical properties of the material.~\cite{Petch1953,Hall1954,Salvalaglio2024} 
 Topological defects are nucleated in real materials  when crystallization occurs on experimental time scales since the system quenches from the disordered high-temperature phase into the ordered low-temperature state.~\cite{Rudolph2007,Reichhardt2017}  In particular, in the case of soft condensed matter  these defects are unavoidably nucleated if the order is also frustrated by boundary conditions and/or the disorder of the host media.~\cite{Chaikin,Fasano2002,Peeters2003,Baus2008,VanderBeek2012,Llorens2020b,Aragon2023,Puig2024} For instance, the former is particularly relevant in the case of nanocrystalline matter since the edge of the system induces strong confinement effects.~\cite{Peeters2003,Berdiyorov2003,Grigorieva2006,Zhao2008,Turlo2017,Gonzalez2020,Haridy2026}

Finite assemblies of interacting particles confined to disk-like geometries—such as colloids in circular traps~\cite{bubeck1999} and Wigner molecules in quantum dots~\cite{Reiman2002}—exhibit universal features, including shell formation, boundary-induced defects, and a crossover toward bulk crystalline order. In this context, vortex nanocrystals provide a tunable and experimentally accessible platform to investigate these generic aspects of crystallization under confinement. Vortex matter in type-II superconductors confined to micron-sized samples therefore constitutes an ideal model system to elucidate the role of topological defects during the crystallization of finite assemblies.~\cite{Moshchalkov1995,Geim1997,Palacios1998,Schweigert1998,Wang2001,Dolz2015} The vortex density is readily tuned through the applied magnetic induction, with the average intervortex spacing scaling as $a \propto \sqrt{B}$. In the magnetic-field range considered here, reducing the intervortex spacing enhances the elastic rigidity of the vortex lattice by increasing the energetic cost of lattice distortions and dislocation formation. Consequently, higher magnetic fields correspond to progressively stiffer vortex nanocrystals,~\cite{Blatter1994} whereas lower fields produce softer lattices that are more susceptible to plastic deformation and defect generation. Quenched disorder, originating from material defects that locally pin vortices, competes with the elastic energy of the lattice and promotes defect formation during the crystallization process.~\cite{Puig2022,Puig2024} Finally, confinement effects can be independently controlled through the sample size and geometry.~\cite{Schweigert1998,Wang2002,CejasBolecek2015} This independent tunability of elastic rigidity, quenched disorder, and geometric confinement makes vortex nanocrystals an excellent model system for studying crystallization dynamics in confined systems.

In nanocrystals a depletion of the total binding energy of the system occurs when decreasing the size  down to few thousand  components due to an enhancement of the  surface-to-volume ratio in the number of components. For vortex nanocrystals, this reduction on the number of components produces a depletion of the entropy-jump in the first-order solidification transition.~\cite{Dolz2015} In addition, the structural order of vortex nanocrystals is worsened with respect to  macroscopic vortex crystals.~\cite{CejasBolecek2017} Direct imaging with individual vortex resolution reveals a sudden increase in the density of topological defects near the edges of the samples. This excess of topological defects in vortex nanocrystals induced by confinement might be at the origin of the depletion of the entropy-jump entailed at the first-order solidification transition.~\cite{CejasBolecek2017}  Close to the sample edge vortex rows have a tendency to bend, an effect that heals on going towards the core of the nanocrystal where an ordered  phase is nucleated. This effect is more pronounced on decreasing the size of the nanocrystal and on decreasing the inter-vortex interaction by lowering $B$. All these observations have been revealed in static experimental conditions after following a field-cooling quenching history from the liquid vortex phase down the solid quasi-crystalline Bragg glass phase.

Controlling the physical properties of the vortex nanocrystals resulting from cooling from a high temperature phase requires to gain insight on the dynamics of the formation of the belt of topological defects crowding the edge of the sample and on the mostly isolated dislocations populating the core of the nanocrystal. Here we get insight on this issue by performing Langevin dynamics simulations to study the crystallization of vortex nanocrystals while field-cooling in micron-sized samples with weak point disorder distributed at random. The choice of cooling protocol, disorder type, and sample geometry  considered in the simulations is driven by experimental results revealing snapshots of nanocrystals of few thousands of vortices crystallized in micron-sized Bi$_{2}$Sr$_{2}$CaCu$_{2}$O$_{8 + \delta}$ samples after following a field-cooling protocol at low applied fields.~\cite{CejasBolecek2015,CejasBolecek2017} As already known from data in macroscopic samples, the vortex structure observed  at the end of the cooling protocol corresponds to a configuration frozen at a characteristic temperature  located below  the melting temperature.~\cite{Dolz2014,CejasBolecek2015}  At this temperature  the dynamics of the vortex system  at length scales of the lattice spacing is appreciably slowed down due to the viscous effect of the unavoidable disorder present in real samples.  
We employ an effective two-dimensional model of interacting vortices to describe the quenched configurations observed in magnetic decoration experiments on Bi$_{2}$Sr$_{2}$CaCu$_{2}$O$_{8 + \delta}$. This description is phenomenological and not intended as a microscopic model of the full three-dimensional vortex lines.

Using this model we present a systematic study of the dynamics of  the formation of the belt of topological defects crowding the edge of the sample and of the mostly isolated dislocations populating the core of the nanocrystal, as a function of the number of vortices in the nanocrystal. The number of vortices is tuned either by the  physical size of the disk-like samples with diameter $D$ or by the vortex density controlled by the magnetic induction $B$. Changes in $B$ also modify the elasticity of the structure: The smaller (larger) $B$ the softer (stiffer) the vortex nanocrystal.  We find that this phenomenological study quantitatively describes the experimental data available in Bi$_{2}$Sr$_{2}$CaCu$_{2}$O$_{8 + \delta}$ nanocrystals. In particular, it allows us to disentangle the intrinsic effect of elasticity and disorder from the effects of confinement controlled by the perimeter-to-area ratio. We find that the dynamics of crystallization presents a characteristic crossover freezing temperature, smaller than the melting one, below which the radial density of topological defects reaches a stationary profile. This freezing temperature is sensitive to both, intrinsic and confinement effects: Increases on raising field and the perimeter-to-area ratio.
The stationary profile of defects is characterized by a healing length at the edge of the nanocrystal that quantitatively agrees with experimental observations. Also in accordance with experiments, the density of defects at the core of the nanocrystal after the field-cooling crystallization is mainly governed by the elasticity of the structure and the disorder of the host media. In contrast, we find that the characteristic healing length in units of $a$ depends mainly on $D$, increasing when the perimeter-to-area ratio decreases. Namely, the number of lattice spacings at the edge of the nanocrystal where the excess of defects is healed is independent of the elasticity of the vortex structure.

\section*{Methods}

We perform two-dimensional Langevin dynamics simulations  mimicking the interaction between rigid three-dimensional vortices~\cite{Reihhardt1997} 
in micron-sized samples with weak point disorder. We focus on the crystallization of vortex nanocrystals during a field-cooling process from a high-temperature disordered vortex structure to a low-temperature state, for various vortex densities in disk-like thin samples with diameters $D$. For selected temperatures while cooling, we quantify the proliferation of topological defects, mostly dislocations, by considering their radial density $\rho(r) = N_{\rm def}(r)/N_{\rm v}(r)$. Here $N_{\rm def}(r)$ is the number of non-sixfold coordinated vortices and $N_{\rm v}(r)$ the total number of  vortices in a circular shell  with radius $r$ centered at the sample center.
We denote $\rho_{\rm T}$ as the total density of defects in the nanocrystal at a given temperature $T$.

For each vortex located at $\mathbf{r}_i$, with neighbors at $\mathbf{r}_j$, the equation of motion for its overdamped Langevin dynamics reads 
\begin{eqnarray}
      \eta \frac{\partial \mathbf{r}_i}{\partial t} &= 
       \sum_{j\neq i}   
      \frac{\epsilon_0}{\lambda_{\rm ab}}
      K_1\left(\frac{|{\bf r}_i-{\bf r}_j|}{\lambda_{\rm ab}}\right)
      \frac{{\bf r}_i-{\bf r}_j}{|{\bf r}_i-{\bf r}_j|} \nonumber \\
          &- \nabla V_{\rm p}(\mathbf{r}_i)
     - \nabla V_{\rm d}(\mathbf{r}_i)
     + \mathbf{f}_i(t)  
\label{eq:equationofmotion}
\end{eqnarray}

\noindent 
where $\eta$ is an effective viscosity proportional to the Bardeen-Stephen viscosity, $\epsilon_{0}$ is the vortex line tension, $\lambda_{\rm ab}$ the superconducting penetration depth and $K_{1}$ is the first-order modified Bessel function. The term $\mathbf{f}_i$ is a random force accounting for thermal fluctuations and satisfying 
$\langle \mathbf{f}_i(t)\rangle =0$ and $\langle \mathbf{f}_i(t) \mathbf{f}_j(t') \rangle = 2 \eta k_B T \delta_{ij} \delta(t-t')$
with $\langle \dots \rangle$  an ensemble average. 
We consider the pinning potential

\begin{equation}
    \begin{aligned}
V_{\rm p}(\mathbf{r}) = -\sum_{p=1}^{N_{p}}  V_{\rm p} \exp{[-|\mathbf{r} - \mathbf{r}_{\rm p}|^2/2\zeta^2]}
\end{aligned}
    \label{eq:Vpinning}
\end{equation}

\noindent that emulates the disorder generated by a random and dense spatial distribution of weak point pins located at $r_{\rm p}$, all with equal magnitude $V_{\rm p}$, and present in the samples with a density $\rho_{\rm p}=N_{p}/A$ in the simulation area $A$. The magnitude $\zeta$ is the range of the point pinning potential that we consider equal to $0.1\lambda_{\rm ab}$ for all the simulations. 
The confinement potential is modeled as a truncated parabola

\begin{align}
    V_d(\mathbf{r})=\frac{\epsilon_0}{2\lambda_{ab}^2}  r^2 \Theta(\lambda_{ab}-|R \hat{r}-\mathbf{r}|)
\end{align}
with $\mathbf{r}=0$ the center of the disk of radius $R = D/2$. The equations of motion \eqref{eq:equationofmotion} are integrated using an explicit Euler scheme, and the computation of the short-range vortex–vortex and vortex–pin interactions is optimized using a double cell-list method. \cite{AllenTildesley2017}

\begin{figure}
\includegraphics[width=0.5 \textwidth,angle=0]{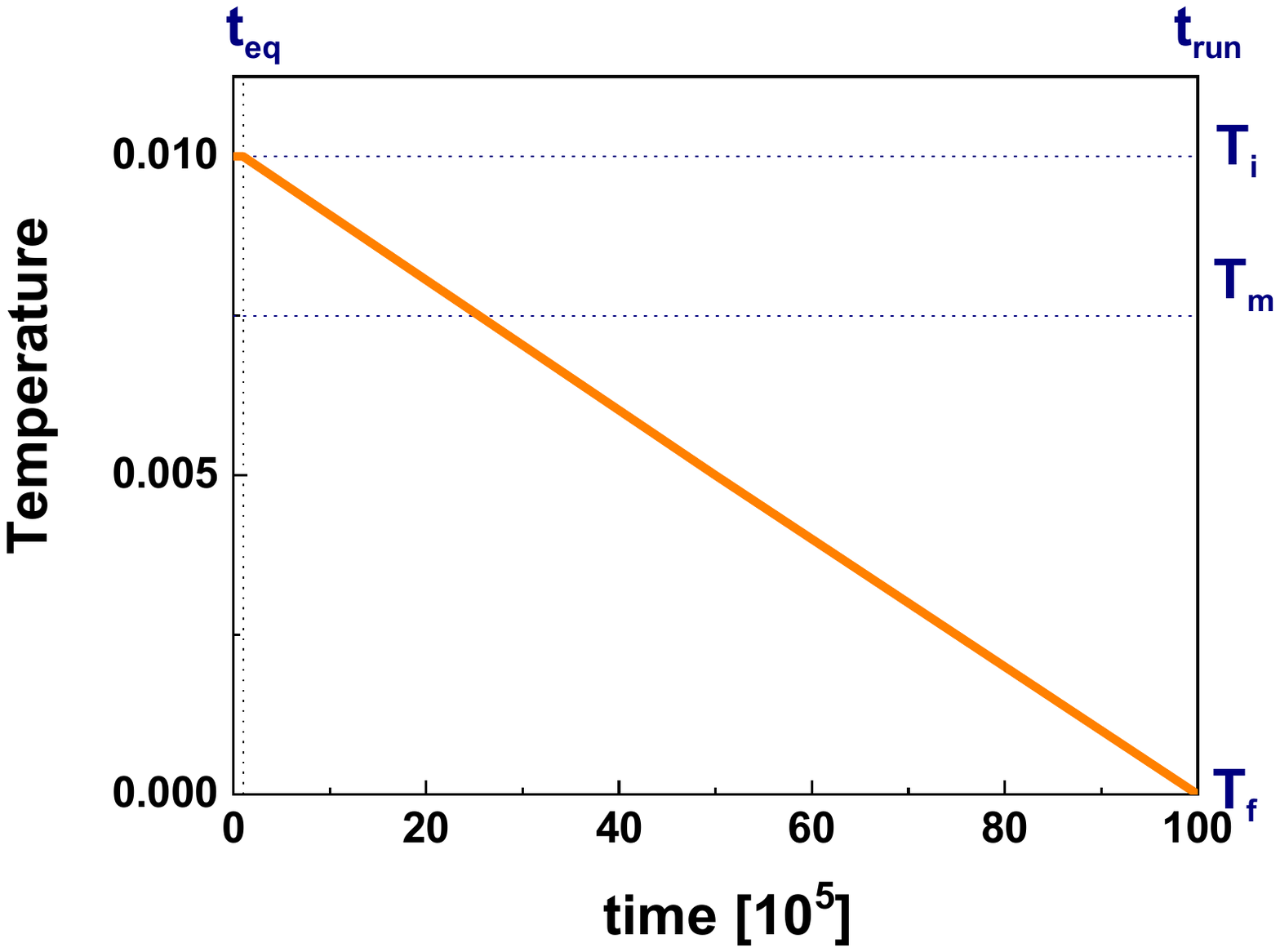}
\caption{Typical temperature-history protocol used in our simulations emulating field-cooling experimental conditions followed in magnetic decoration experiments. The equilibration and ramp times, $t_{\rm eq}$ and $t_{\rm run}$, as well as the initial $T_{\rm i}$, final $T_{\rm f}$, and melting $T_{\rm m}$ temperatures are indicated.
\label{Figure1}}
\end{figure}

We perform  simulations for disks with diameter $D=30$, 40 and 50 $\mu$m and for vortex densities of 16, 32 and 50\,G. The vortex density is controlled by the average separation between vortices, $a = 1.075 \sqrt{\Phi_{0}/B}$ with $\Phi_{0}=2.07 \cdot 10^{-7}$\,G$\cdot$cm$^{2}$ the flux quantum.
The pinning parameters $\rho_p$, $V_p$ and $\xi$, are chosen as to obtain in our simulations a density of topological defects in macroscopic samples similar to those observed in experiments.~\cite{Dolz2015,CejasBolecek2015,CejasBolecek2017} We consider the value of $\lambda_{\rm ab}$ as equal to that of Bi$_{2}$Sr$_{2}$CaCu$_{2}$O$_{8 + \delta}$ at the irreversibility temperature.~\cite{Fasano2005,Dolz2015} All magnitudes in the simulations are dimensionless since we measure distance in units of $\lambda_{\rm ab}$, time in units
of $\eta \lambda_{\rm ab}^{2}/\epsilon_{0}$, energy per unit length in units of $\epsilon_{0}\lambda_{\rm ab}$, and temperature  in
units of $\epsilon_{0}\lambda_{\rm ab}/k_{\rm B}$ per unit length, with $k_{\rm B}$ the Boltzmann constant. In what follows we
refer exclusively to dimensionless variables.

We simulate field-cooling protocols resembling the thermal history followed in experiments. 
We consider as the initial condition at $T_{\rm i}=0.01$ a
thermally-equilibrated 
vortex liquid configuration confined in the disks.
The system is then
cooled down to $T_{\rm f}=0.0001$ following a linear decrease of temperature during a ramp time $t_{\rm run}$.   Figure \ref{Figure1} shows a schematic representation of one particular cooling protocol performed in the simulations.
We vary the sweep rate of the temperature ramp, $v_{\rm sweep} = (T_{\rm i} - T_{\rm f})/t_{\rm run}$, by changing the value of $t_{\rm run}$ in the range of $10^{3}$ to $10^{7}$ simulation steps for the fastest and slowest ramps, respectively. As discussed in the next section, during this temperature-ramps the vortex nanocrystals transform at the melting temperature $T_{\rm m}$ from a liquid at high temperatures to a solid state at $T_{\rm f}$ .

\begin{figure}
\includegraphics[width=\columnwidth,angle=0]{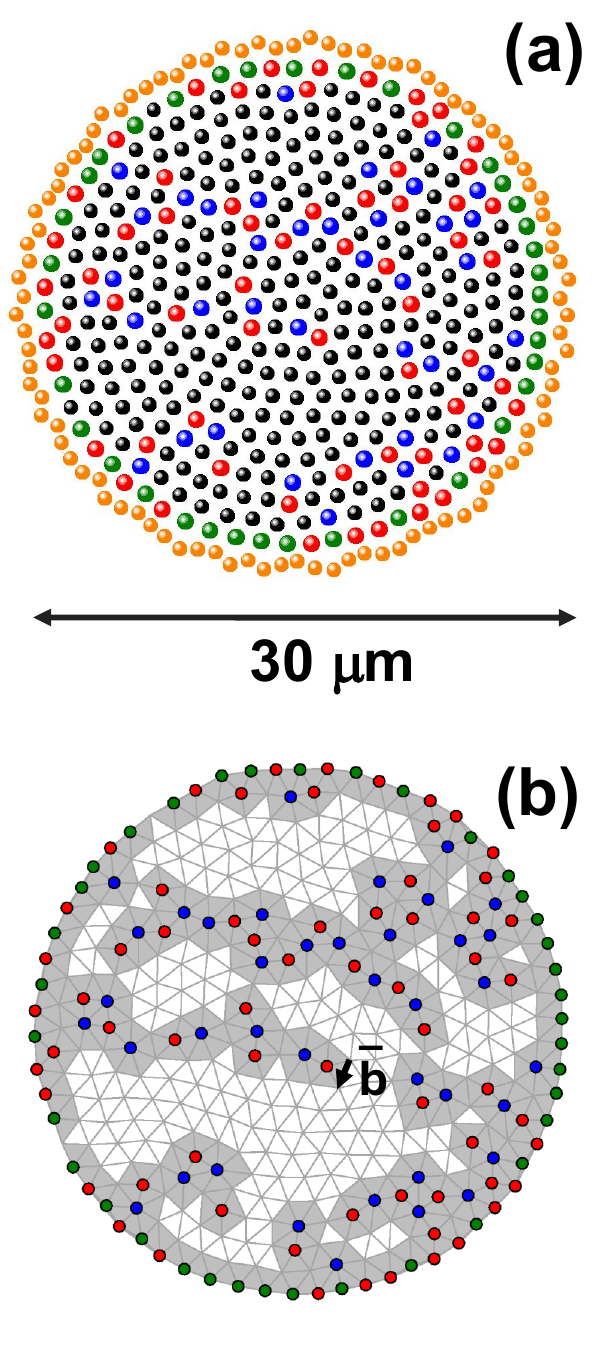}
\caption{(a) Example of a snapshot of a nanocrystal with $\sim 400$ vortices crystallized at $T_{\rm f}$ after the field-cooling protocol followed in our simulations. Results correspond to the smallest magnetic field studied of 16\,G and for a disk with 30\,$\mu$m diameter. Non-sixfold-coordinated vortices are 
color-coded for 7 (blue), 5 (red), 4 (green) and 8 (yellow) neighbors.  The orange vortices  at the edge of the nanocrystal are not considered for the calculation of the defect densities presented in this work. (b) Delaunay triangulation of the vortex nanocrystal with topological defects highlighted in gray.  The Burgers vector of one isolated dislocation, $\mathbf{b}$, is indicated. 
\label{Figure2}}
\end{figure}

For each studied vortex density $B$ and sample size $D$ we perform five independent simulations of field-cooling processes starting at $T_{\rm i}$, finishing at $T_{\rm f}$, and following the same sweep rate while cooling. The five independent simulations are performed by starting with five different initial vortex distributions at high temperature, with different Langevin noise realizations.  All the magnitudes reported in the next section are averaged over these 5 simulation runs and the error bars in the data represent the standard deviation in this average, achieving a statistical precision better than 10\%.



\section*{Results}

We study the dynamics of topological defects in nanocrystals while cooling by, computing the density of topological defects as a function of temperature, $\rho_{\rm T}$. We study different vortex densities and sample sizes comparable to available experimental data.~\cite{CejasBolecek2017} In order to obtain $\rho$ we start considering snapshots of the nanocrystals as for instance the one shown in Fig.\,\ref{Figure2} (a) for the smallest nanocrystal studied. By means of a Delaunay triangulation algorithm~\cite{Fasano2003} we identify sixfold and non-sixfold coordinated vortices,  the latter color-coded as indicated in the figure. The density of defects is computed inside the nanocrystal, without considering the vortices located at its very edge, see  orange  vortices in Fig.\,\ref{Figure2} (a). $\rho$ is estimated as the ratio between the non-sixfold to the total number of vortices.

Inside of the nanocrystal, most of non-sixfold coordinated vortices have 5 and 7 neighbors, topological defects known as disclinations. They are generally paired forming edge dislocations, see gray-highlighted triangles in the Delaunay triangulation of Fig.\,\ref{Figure2} (b). The region close to the edge of the nanocrystal is highly crowded with dislocations since the vortex arrangement close to the edge mimics the circular shape of the disk. On going towards the center of the nanocrystal most dislocations are located in grain boundaries separating small crystallites with different orientations. Isolated dislocations are not excesively frequent,  particularly in the smallest nanocrystals, though they generally appear close to the center of the sample. The topological charge of dislocations known as Burgers vectors,~\cite{Hull2011} see for example arrow in Fig.\,\ref{Figure2} (b), are related to the number of extra planes of vortices crossed when encircling the dislocation. Some pairs of Burgers vectors annihilate in close-by dislocations and some pairs annihilate in far apart dislocations.

\begin{figure}
\hspace{-1 cm}
\includegraphics[width=\columnwidth,angle=0]{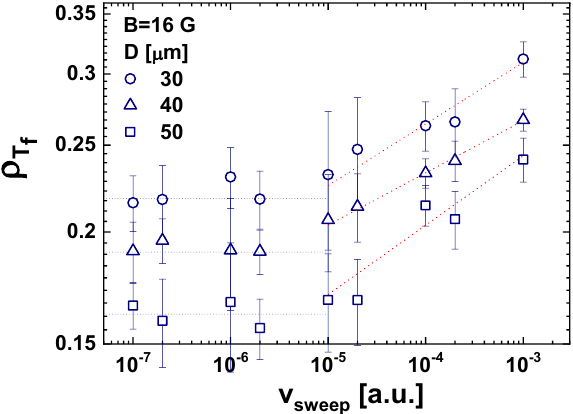}
\caption{Total density of defects in the vortex nanocrystal at the final temperature, $\rho_{T_{\rm f}}$, as a function of the temperature sweep-rate  followed while cooling, $v_{\rm sweep}$. Data for a vortex density of 16\,G and disk samples with diameters $D=30$, 40 and 50\,$\mu$m.  Blue dotted lines are fits to the data with a constant for the slow $v_{\rm sweep}$ regime whereas red dotted lines are algebraic growth fits for the fast sweep-rates. }
\label{Figure3}
\end{figure}

 We first start by studying how changes in the temperature sweep-rate of the simulations alter the total density of defects at the final temperature, $\rho_{\rm T_{\rm f}}$. Figure\,\ref{Figure3} shows $\rho_{T_{\rm f}}$ for simulations performed at various sweep-rates, $v_{\rm sweep}$, at a fixed vortex density of 16\,G for disks with 30, 40 and 50\,$\mu$m diameters. $\rho_{T_{\rm f}}$ enhances with increasing $v_{\rm sweep}$ presenting two characteristic behaviors within the error of the data: For large sweep-rates grows algebraically whereas for  $v_{\rm sweep}< 10^{-5}$ display a tendency to saturation.  We observe this phenomenology when performing simulations spanning changes in the cooling rate of up to four orders of magnitude.  The $\rho_{T_{\rm f}}$ presents a clear tendency to saturation for the four slowest cooling rates studied. A similar phenomenology is observed for the other vortex densities studied in this work. Thus, in order to avoid spurious effects in the total density of defects due to  fast cooling ramps, in what follows we perform slow simulations considering $v_{\rm sweep}=10^{-7}$.

\begin{figure*}
\includegraphics[width=2\columnwidth,angle=0]{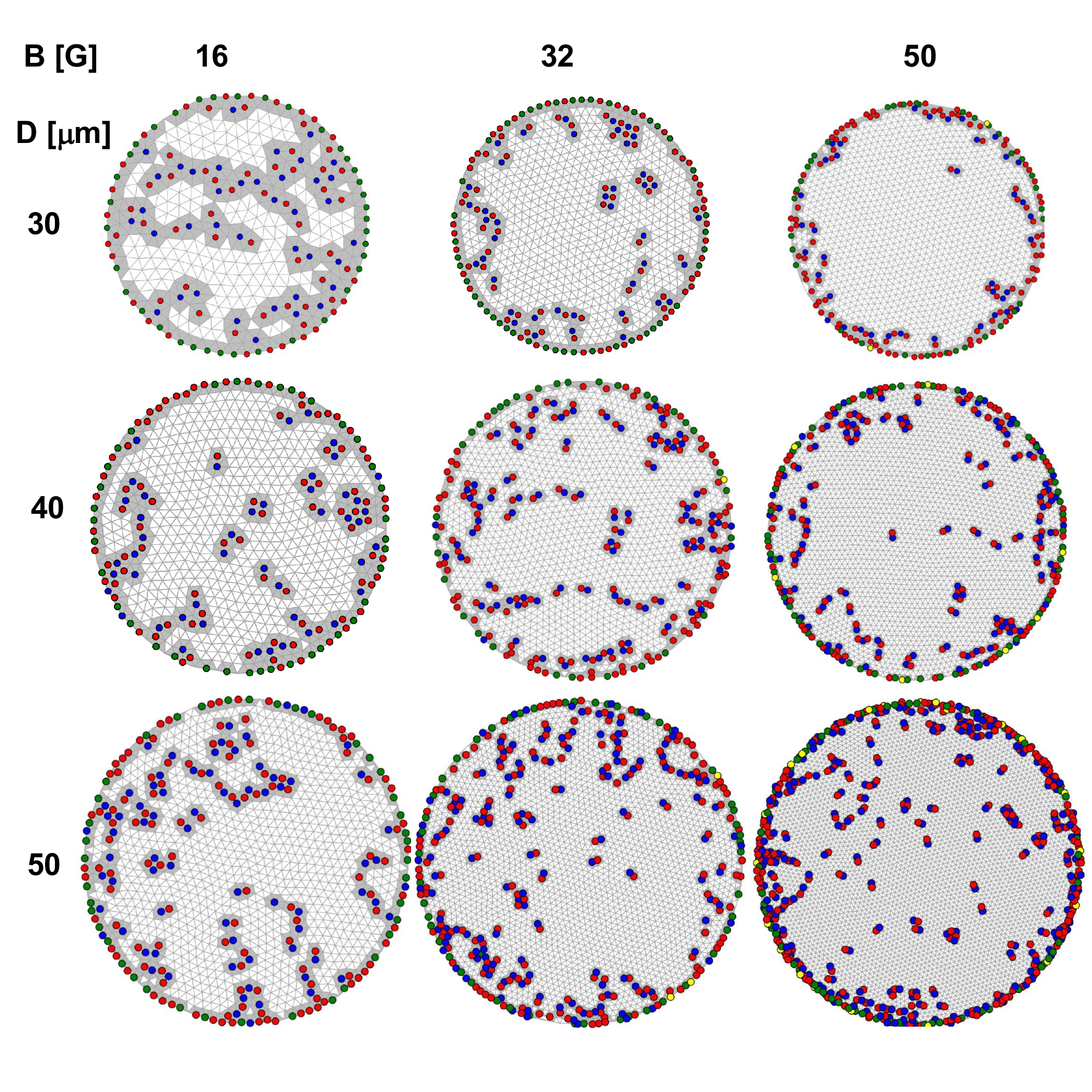}
\caption{Examples of Delaunay triangulations of the snapshots of vortex nanocrystals obtained at the lowest temperature, $T_{\rm f}$, after following field-cooling protocols at the slowest sweep-rate of $v_{\rm sweep} = 10^{-7}$. Columns present data at fixed vortex densities of 16, 32 and 50\,G whereas rows show data at fixed sample sizes with diameters $D=30, 40, 50$\,$\mu$m.  Topological defects are highlighted in gray and non-sixfold-coordinated vortices are 
color-coded for 7 (blue), 5 (red), 4 (green) and 8 (yellow) neighbors. Vortices at the edge of the nanocrystal are not considered for the calculation of the Delaunay triangulation and defect density (not shown in the figure).
\label{Figure4a}}
\end{figure*}

\begin{figure*}
\includegraphics[width=2\columnwidth,angle=0]{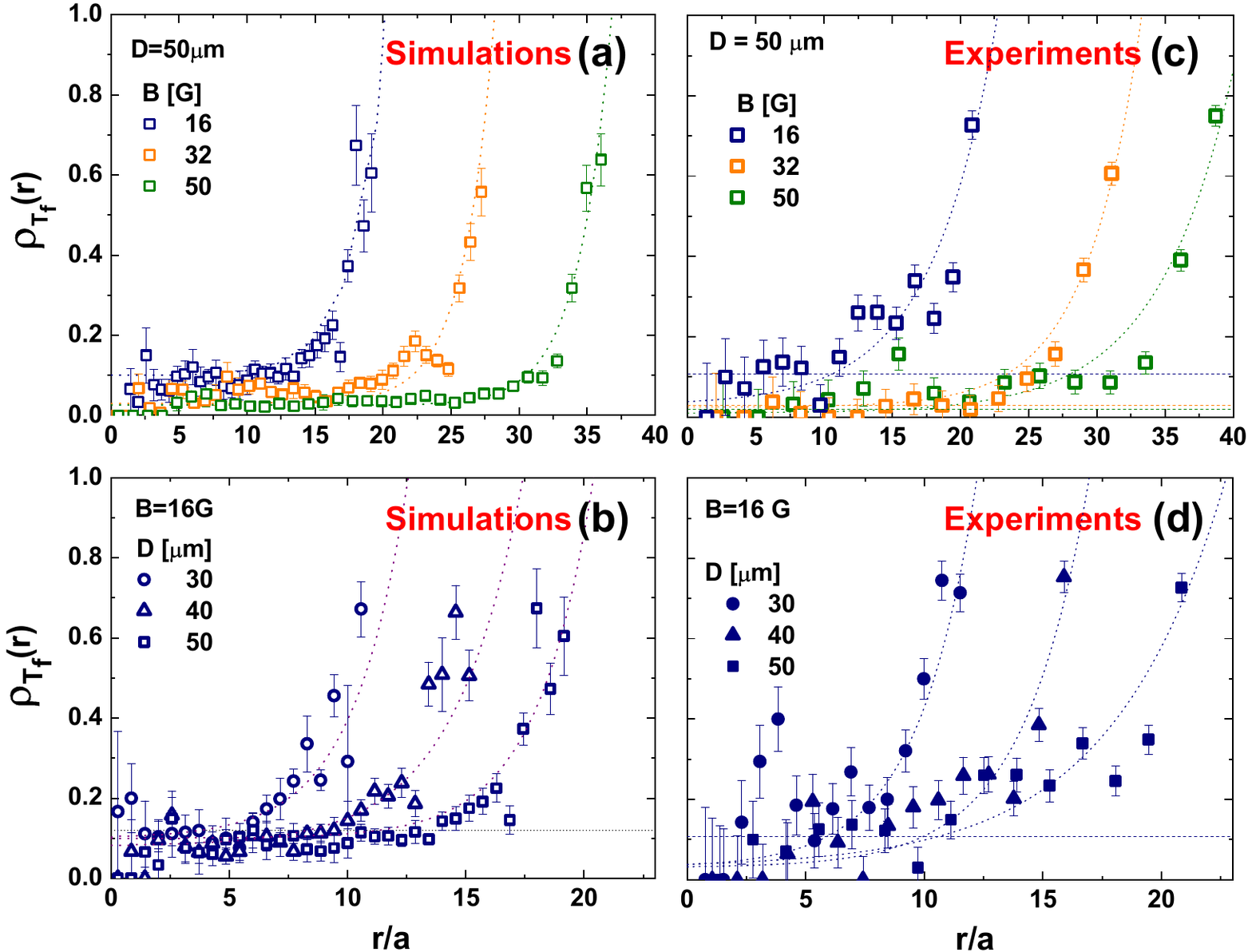}
\caption{ Simulated and experimental radial density of defects in the vortex nanocrystals at the final temperature, $\rho_{T_{\rm f}}(r)$ ($T_{\rm f} =0.0001$ for simulations and 4.2\,K for experiments), as a function of the radius normalized by the average vortex separation, $r/a$.  (a) Simulation data for a given diameter $D$ of nanocrystals with different number of vortices between 1300 and 4700 tuned by changing the induction $B$ in the range between 16 and 50\,G. (b) Simulation data for a vortex density of 16\,G and disk samples with diameters $D=30$, 40 and 50\,$\mu$m. (c,d) Experimental data from Ref.\,\onlinecite{CejasBolecek2017} for the same values of $D$ and fields shown in panels (a) and (b). Dotted lines are fits to the data with a function that considers a saturation value plus an exponential growth on going towards the edge of the nanocrystals. For the simulation data the outer shell of vortices is discarded as indicated in Fig.\,\ref{Figure2} whereas in the case of the experimental data taken from Ref.\,\onlinecite{CejasBolecek2017} the vortices from the outer shell are considered. 
\label{Figure4}}
\end{figure*}

Figure\,\ref{Figure4a} shows examples of the simulated vortex positions obtained at $T_{\rm f}$ after following a field-cooling protocol at the slowest sweep-rate. This figure presents  Delaunay triangulations highlighting the topological defects for vortex densities of 16, 32 and 50\,G, and samples with diameters $D=30, 40, 50$\,$\mu$m. These images show that  for a fixed vortex density, confinement effects enhance on decreasing the size of the nanocrystals, see for instance the saturation values of $\rho_{\rm T_{\rm f}}$ in Fig.\,\ref{Figure3}. Data for samples with a given diameter $D$ and different number of vortices controlled by  $B$ indicate that, at a fixed $r/a$ value,  the radial density of defects $\rho_{T_{\rm f}}(r)$ is larger for nanocrystals with a lesser number of vortices. For example, Fig.\,\ref{Figure4} (a) shows the case of $D=50$\,$\mu$m and vortex nanocrystals with 1300 (16\,G), 2800 (32\,G), and 4700 (50\,G) vortices. This radial density is obtained by computing the ratio between the number of non-sixfold coordinated vortices and the total number of vortices located in a circular belt of radius $r$ and width $\Delta r = \pm 0.5 a$.  Figure\,\ref{Figure4} (b) also puts in evidence this confinement effect from another set of data:  for a given distance $r/a$ from the center of the nanocrystal $\rho_{T_{\rm f}}(r)$ is larger for smaller disks.

The defect configurations observed during freezing also bear a qualitative resemblance to the grain-boundary scars reported in crystalline order on curved surfaces \cite{bowick2000,bausch2003}. In those systems, however, scars originate from the geometric frustration imposed by Gaussian curvature and the associated topological requirement of excess disclinations. By contrast, the present vortex nanocrystals crystallize on a flat geometry, where the observed defect chains arise from the competition between elastic relaxation, boundary confinement, quenched disorder, and non-equilibrium freezing. Thus, while the underlying mechanisms differ, both systems illustrate how extended dislocation structures provide an efficient route for relieving accumulated elastic stresses.

For all the studied vortex densities and nanocrystal sizes, the density of defects at $T_{\rm f}$ presents a roughly constant value at around the center of the nanocrystal, see horizontal dotted blue line in Fig.\,\ref{Figure4} (b), and enhances  on going towards the edges of the samples. $\rho_{T_{\rm f}}(r)$ remains constant up to a larger $r$ when increasing the diameter of the nanocrystal $D$ for a given vortex density. This saturation value is roughly the same for a given vortex density irrespective of the diameter of the nanocrystal. The enhancement of  $\rho_{T_{\rm f}}(r)$ at the edge of the vortex nanocrystals follows a sudden growth.

This stagnation and growth phenomenology on going towards the edge of the nanocrystals was observed experimentally,~\cite{CejasBolecek2017} and the distance in which the density of defects decreases on going towards the center of the nanocrystal was interpreted as a \textit{healing length}. Moreover, the good quantitative agreement between the simulation and previously published~\cite{CejasBolecek2017} experimental information is clear from the comparison of the data shown in the left and right panels of Fig.\,\ref{Figure4}. In order to describe the observed phenomenology, we follow the lead of the previous experimental work and  fit 

\begin{equation}
\rho_{T_{\rm f}}(r) 
\approx
\rho_{\rm bulk} + A \exp{(r-D/2)/(\alpha\,\cdot\,a)}, 
\end{equation}

\noindent with $\rho_{\rm bulk}$ the  density of topological defects in macroscopic crystals found experimentally for the same vortex density,    and $\alpha\,\cdot\,a$ the healing length. The fits, shown with color dotted lines in the two panels of  Fig.\,\ref{Figure4}, reasonably follow the simulation data within the error.

The phenomenological healing-length parameter $\alpha\,\cdot\,a$
is a measure of the confinement effect introduced by the edge of the sample that bends the lanes of vortices close to the edge by inducing topological defects. Towards the center of the sample, the vortex lattice recovers a hexagonal symmetry with mostly parallel lanes of vortices and the density of topological defects is similar to both, the 
central density of defects observed in experiments in disks and in large macroscopic samples for the same vortex density.  This coincidence is the result of conveniently choosing the magnitude of the point pinning potential parameters  $V_{\rm p}= 0.004$ and $\rho_{p}=25$ in order to quantitatively reproduce the experimental density of defects at the center of the disks that is also roughly equal to the average density of defects in macroscopic samples at a given field. Variations of these parameters in 10\% or more produce changes in the simulated density of defects that deviate between 10-50\% from the values observed experimentally in an extended field range. As mentioned, the range of the point pinning potential $\zeta=0.1\lambda_{\rm ab}$, of the order of the coherence length of the high-temperature superconductor studied experimentally in Ref.\,~\onlinecite{CejasBolecek2017}.

\begin{figure}
\includegraphics[width=0.92\columnwidth,angle=0]{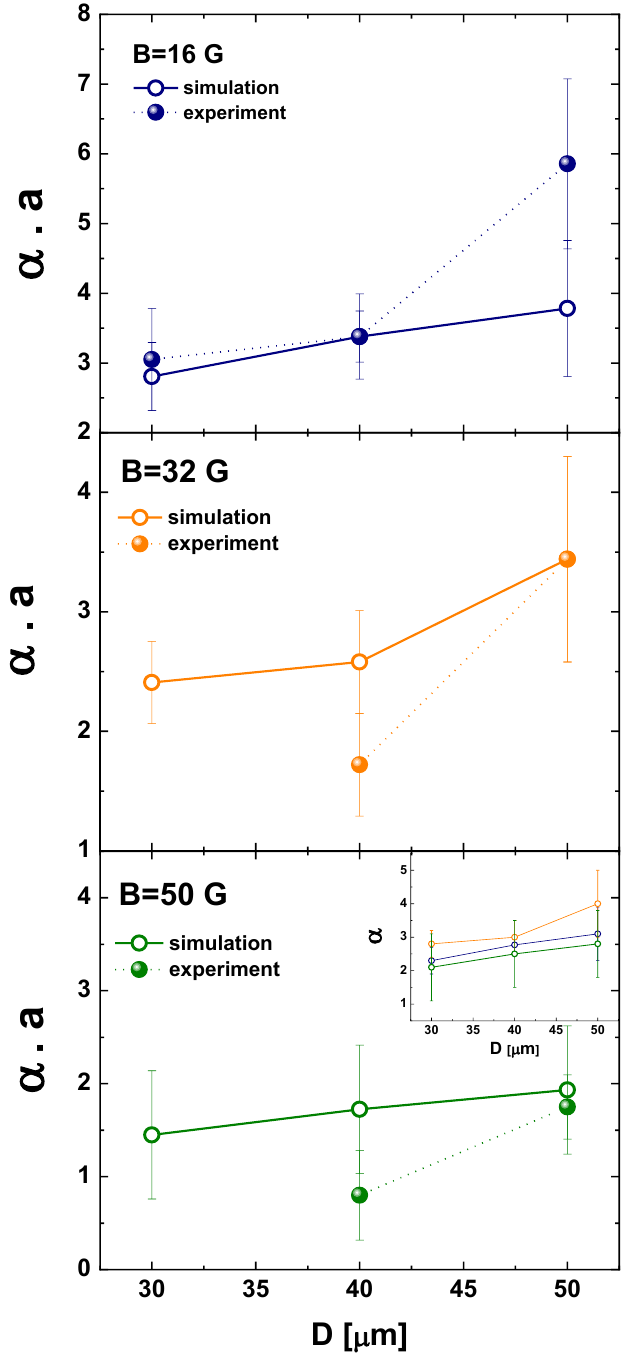}
\caption{Dependence of the phenomenological healing length $\alpha \cdot a$ with the diameter $D$ of the vortex nanocrystal for vortex densities of (a) 16, (b) 32 and (c) 50\,G. Simulation results (open dots) are compared with experimental data (full dots) from Ref.\,\onlinecite{CejasBolecek2017}. Insert: Parameter $\alpha$ obtained in the simulations for all the studied vortex densities and sample sizes.
\label{Figure5}}
\end{figure}

Figure\,\ref{Figure5} presents the variation of $\alpha \cdot a$  with $D$ and shows that the simulation data agrees  quantitatively well with the experiments within the error. Both in simulations and experiments the healing length increases on growing $D$ at a given vortex density.~\cite{CejasBolecek2017} 
In addition, for a given size of the sample, $\alpha$ is almost independent of the vortex density, see insert to Fig.\,\ref{Figure5}. Then, the characteristic healing length in units of $a$ depends primarily on $D$ such that enhances when the perimeter-to-area ratio diminishes.

\begin{figure}
\includegraphics[width=0.95\columnwidth,angle=0]{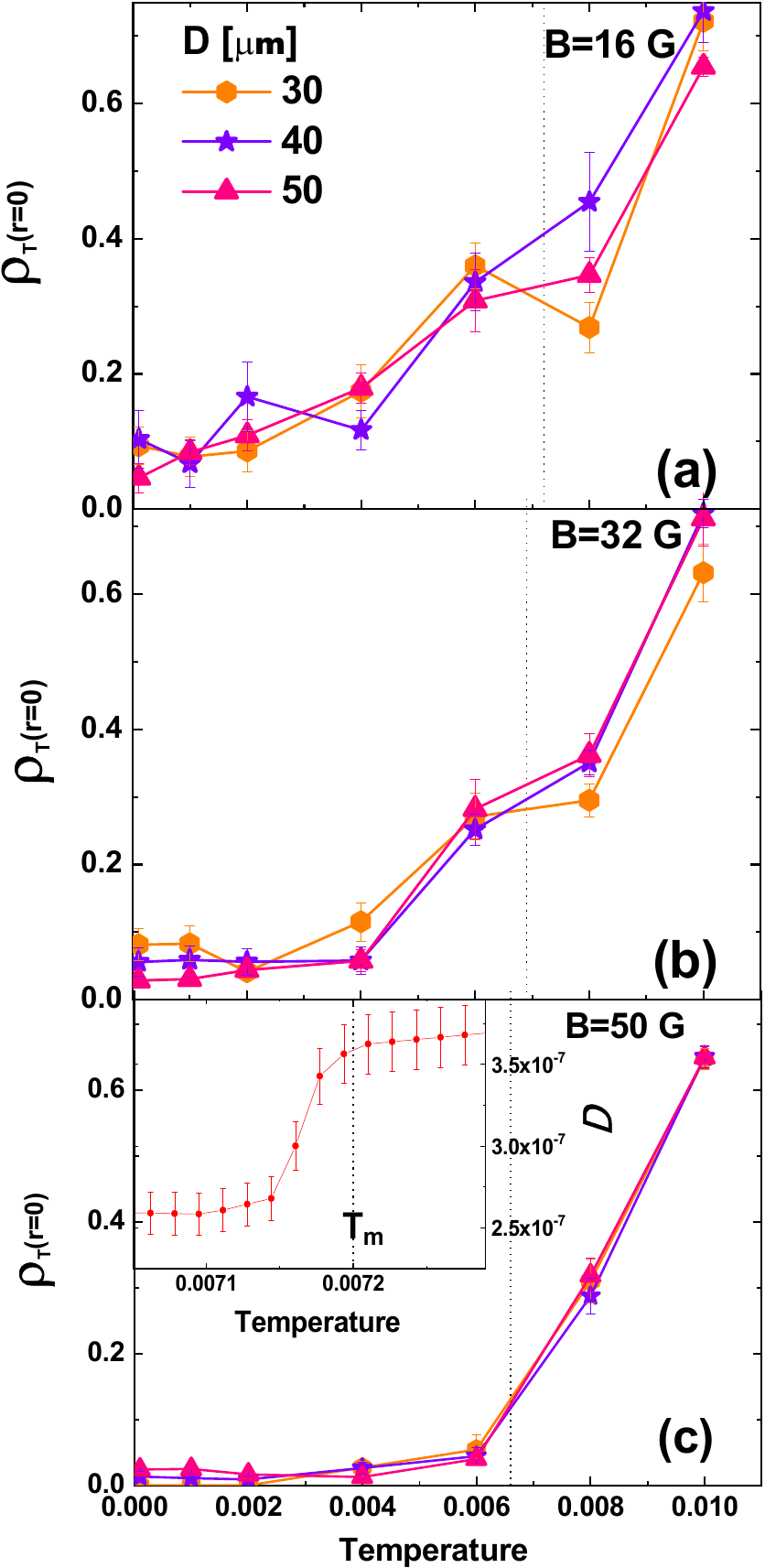}
\caption{Central density of topological defects as a function of temperature, $\rho_{\rm T}(r=0)$ during the field-cooling process starting at $T_{\rm i}=0.01$ and finishing at $T_{\rm f} = 0.0001$. Data for different disk diameters $D$ and vortex densities of (a) 16, (b) 32 and (c) 50\,G. The melting temperature $T_{\rm m}$ at each vortex density is indicated with vertical lines.  Insert: Example of the determination of $T_{\rm m}$  from the abrupt jump in the diffusion coefficient of the vortex structure for the case of $B=16$\,G.
\label{Figure6}}
\end{figure}

In order to study the dynamics of this healing process, we start by considering the density of defects at the center of the sample for a given temperature $T$ while field cooling, namely $\rho_{\rm T}(r=0)$. Figure\,\ref{Figure6} presents the variation with temperature of the average of this magnitude over 5 independent runs for all the simulated vortex densities and sample sizes. For every vortex density studied, the curves for different sample diameters roughly overlap within the error.  Irrespective of vortex density, at $T_{\rm i}$ when the  field cooling starts, $\rho_{\rm T}(r=0) \sim 0.7$, as expected for a very disordered liquid vortex structure. On starting cooling, $\rho_{\rm T}(r=0)$  decreases significantly and at an intermediate temperature displays a tendency to saturation before reaching $T_{\rm f}$. This characteristic temperature depends on vortex density, being lower for softer vortex structures nucleated at smaller $B$. For a given vortex density, no clear dependence of this saturation phenomena with $D$ is detected in our simulations. These phenomenology indicates that the dynamics of crystallization of the nanocrystals gets frozen at an intermediate temperature that depends primarily on the system elastic properties: On increasing the stiffness of the nanocrystal it gets frozen at a higher temperature.

This freezing temperature is, in all cases studied, smaller than the field-dependent melting temperature of the nanocrystals, $T_{\rm m}$, see dotted vertical lines in Fig.\,\ref{Figure6}. The insert to this figure shows an example on how we determine the melting transition in our simulations from the sudden jump on the diffusion coefficient $\mathcal{D}$ of vortices on cooling.   Data on the temperature-dependent $\mathcal{D}(T)$ are obtained from 
a different set of simulations where the vortex structure is cooled at the slowest sweep rate down to a given temperature $T$ at which the system is allowed to evolve for a long time $t_{0}$. After that time we compute the mean-square displacement of vortex positions that is proportional to $t-t_{0}$ via the difussion coefficient, namely $\Delta^{2}(T)/a^{2} = \mathcal{D}(T) (t-t_{0})$ for sufficiently large 
$t_0$ and $t-t_0$, according to the Einstein-Smolochuwski relation.~\cite{CruzGarcia2026} Here we consider the melting temperature as that where $\mathcal{D}$ starts to register a sudden jump on cooling, see vertical dotted line in the insert to Fig.\,\ref{Figure6}.

\begin{figure}
\includegraphics[width=1.1\columnwidth,angle=0]{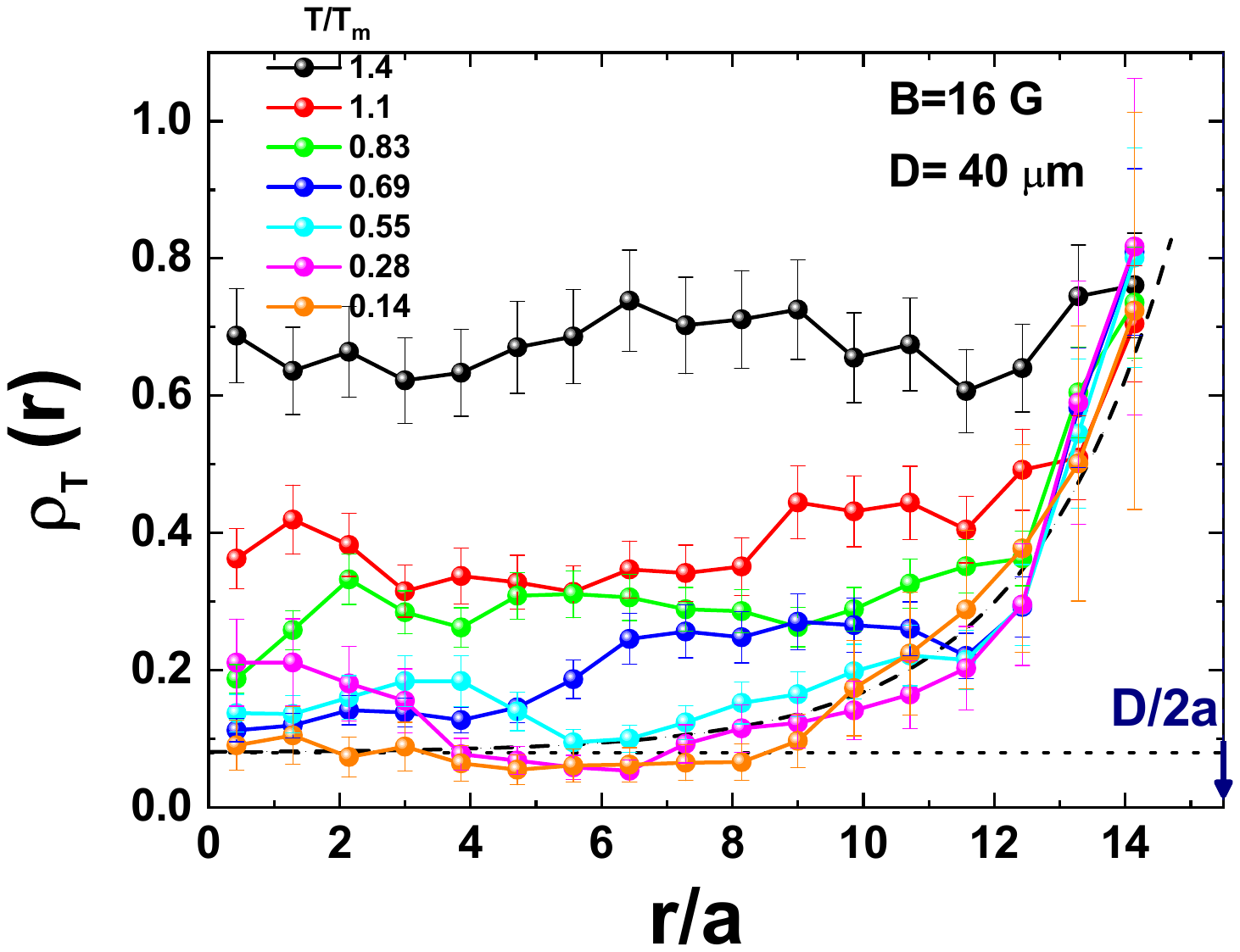}
\caption{Radial density of topological defects as a function of distance from the center of the nanocrystal, $\rho_{\rm T}(r)$ for various temperatures while field-cooling through the vortex melting transition at $T_{\rm m}$. Data for a disk with $D=40$\,$\mu$m and a vortex density of 16\,G. The dashed black line is a fit to the data at the lowest temperature of $T/T_{\rm m}=0.14$ with a function $\rho_{\rm bulk} + A \exp{(r-D/2)/(\alpha\,\cdot\,a)}$,  with $\rho_{\rm bulk}$ the  density of topological defects in macroscopic crystals found experimentally for the same vortex density  (see dotted horizontal line).
\label{Figure7}}
\end{figure}

The saturation of the central defect density $\rho_{\rm T}(r=0)$ at an intermediate temperature smaller than $T_{\rm m}$ is also consistent with the evolution with temperature of the radial density of defects for a given temperature $T$, $\rho_{\rm T}(r)$. Figure\,\ref{Figure7} shows for example the changes on this radial magnitude  on cooling  a vortex nanocrystal with a density of 16\,G nucleated in a sample with $D=40$\,$\mu$m. As in all the studied cases, the density of defects at the center of the sample decreases on cooling and stagnates at an intermediate temperature, in this case around $0.7 T_{\rm m}$. In addition, all the curves at different temperatures depict the mentioned healing effect on going towards the center of the sample, effect that becomes steeper on decreasing temperature.  In this example, for the three smallest temperatures presented, the $\rho_{\rm T}(r)$ curves roughly overlap in most of the $r/a$ range, suggesting the dynamics of crystallization and thus the healing process of the nanocrystal is already frozen at all lengthscales. A similar qualitative behavior on the temperature evolution of the radial density of defects is observed for all studied cases.

The freezing process spans in  a finite temperature range below the melting transition and as we discuss now it is a crossover process that occurs at different temperatures in different locations of the sample. In order to show this, consider for instance the case of the vortex structure nucleated at 16\,G in a $40$\,$\mu$m diameter disk. Panel (a) of Fig.\,\ref{Figure6} indicates that at the center of the sample the density of topological defects is still significantly varying  at $T=0.004 \sim 0.55 T_{\rm m}$. In contrast, Fig.\,\ref{Figure7} shows that at this temperature, for $r/a > 12$ the density of defects is similar to values found at lower temperatures. Similarly, for $T=0.005 \sim 0.69 T_{\rm m}$,
$\rho_{\rm T}(r)$ at distances $r/a < 12$ is larger than for smaller temperatures  but register similar values for  $r/a > 12$. This phenomenology suggests that the freezing process is still underway at the center of the sample whereas at its outskirt the nanocrystal gets frozen at higher temperatures in a highly disordered structure.  This temperature-delay in the freezing process when going towards the center of the sample has been observed in all the studied cases.

\begin{figure}
\includegraphics[width=0.95\columnwidth,angle=0]{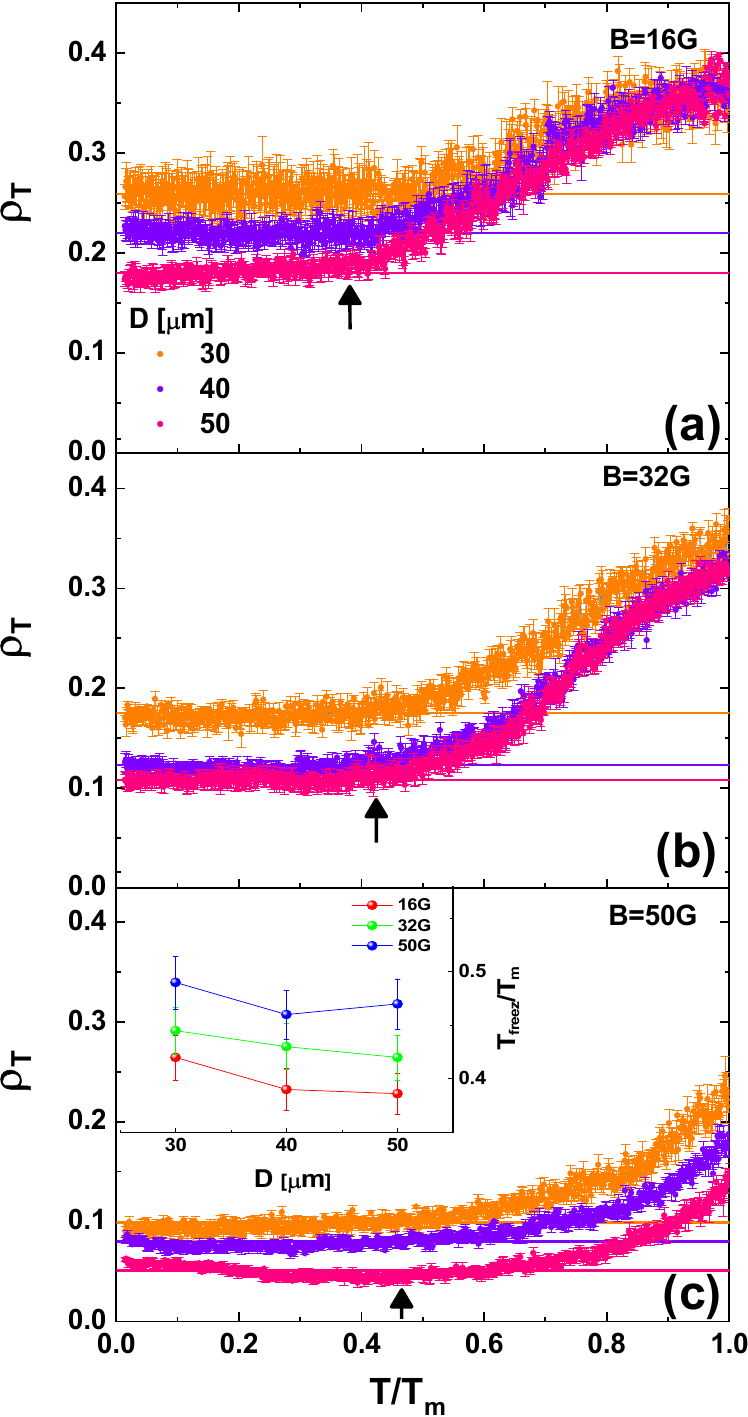}
\caption{Variation with temperature of the total density of defects $\rho_{\rm T}$ while field cooling vortex nanocrystals with densities of (a) $16$, (b) 32 and (c) 50\,G. Data for nanocrystals of different sizes with $D=30, 40, 50$\,$\mu$m. Horizontal lines correspond to the saturation value of the total density of defects once the vortex structure is frozen. The characteristic freezing temperature for the larger disks is indicated with arrows. 
\label{Figure8}}
\end{figure}

\begin{figure*}
\includegraphics[width=2.2\columnwidth,angle=0]{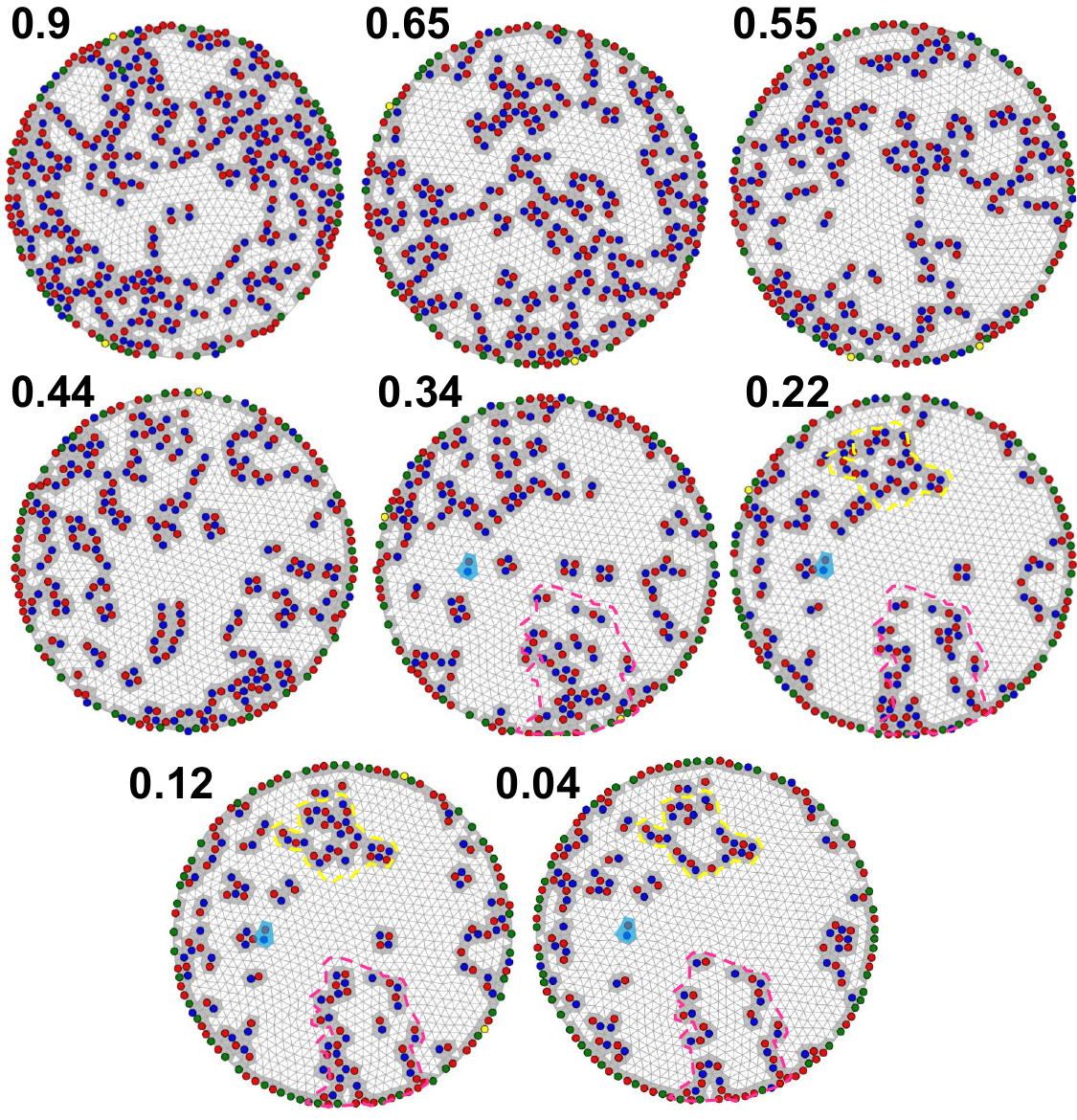}
\caption{Sequence of snapshots while field-cooling a vortex nanocrystal nucleated at 16\,G in a sample with 50\,$\mu$m diameter for the different $T/T_{\rm m}$ temperatures indicated in black. The Delaunay triangulations show sixfold (non-sixfold) coordinated vortices 
color-coded as in the previous figures. Some defects have been highlighted in different colors in order to follow their dynamics on cooling: Pink and yellow crystallites that start to form at intermediate temperatures and are well defined at the lower temperatures, as well as a turquoise dislocation located inside of the nanocrystal remaining rather static from intermediate to low temperatures.
\label{Figure9}}
\end{figure*}

With the aim of following a unique criteria for determining a characteristic freezing temperature for vortex nanocrystals with given size and density, we thus consider a  global magnitude. We recall that it is important to keep in mind that the freezing is a crossover and not an abrupt process that spans in a temperature range rather than at a clearly defined temperature, with this temperature range also depending on the location of the sample. One possible global criteria could be based on the temperature-evolution of the total density of defects in the nanocrystal.  We thus compute the total density of defects in the studied nanocrystals at a given temperature $T$, $\rho_{\rm T}$, and plot it as a function of temperature on cooling, see Fig.\,\ref{Figure8}. In all studied cases $\rho_{\rm T}$ shows a tendency to remain constant  below  a field-dependent temperature, see arrows indicating this temperature for the largest studied disks. We consider this stagnation in $\rho_{\rm T}$
as a global indication that the vortex nanocrystal is frozen and thus consider this temperature $T$ as the freezing temperature $T_{\rm freez}$. The value of $T_{\rm freez}$ 
moderately decreases on increasing $D$, namely on reducing the perimeter-to-area ratio, and has a clear dependence on vortex density being smaller for softer nanocrystals (smaller $B$). This behavior is summarized in the insert to Fig.\,\ref{Figure8} that shows the ratio of $T_{\rm freez}$ to $T_{\rm m}$ as a function of vortex density for all the studied $D$.

Since our simulation results agree  quantitatively well with experimental data,  we feel confident to study the vortex freezing dynamics with this method. 
Experimental field-cooling magnetic decorations do not provide information of nanocrystal freezing dynamics since the observed vortex patterns are snapshots of the structure frozen at a given $T_{\rm freez}$ of the
order or smaller than the temperature at which pinning sets in.~\cite{Fasano1999,Fasano2003} On the contrary, our simulation results can provide valuable information on the freezing dynamics while cooling nanocrystals in micron-sized samples with weak and dense point pinning disorder.

Figure\,\ref{Figure9} shows a sequence of snapshots of the vortex nanocrystal while field-cooling from high to low temperatures $0.04<T/T_{\rm m}<0.9$ for the example of a 16\,G vortex structure nucleated in a disk with 50\,$\mu$m diameter. The sequence highlights the main results already presented: i) A global decrease in the density of defects is produced on cooling; ii) the edge of the sample introduces a confinement effect that tends to bend the structure generating a greater density of topological defects at the outskirt of the nanocrystal; iii) at an intermediate temperature $T_{\rm freez}$ ($\sim (0.44-0.34)T_{\rm m}$ in this case) the nanocrystal gets globally frozen presenting an outer belt of defects and an almost healed ordered structure at the center of the nanocrystal; iv) this freezing process is actually location-dependent since at temperatures slightly below the defined global freezing temperature $T_{\rm freez}$  ($\sim 0.38 T_{\rm m}$ in this case) the defects located at the central part of the nanocrystal still present a certain degree of dynamics, see for instance the dislocation indicated in pink that appears at  $T/T_{\rm m}=0.34$.

\section{Conclusions}

We show that the crystallization of vortex nanocrystals is strongly controlled by the area-to-perimeter ratio, as well as by the elasticity of the system and the disorder in the samples. These factors determine both the total density of defects and its spatial profile at different temperatures, which can be characterized by a healing length and a core value within the nanocrystal. In our study, the profiles of topological defects obtained in low-temperature simulations agree quantitatively well with experimental data, despite we consider a phenomenological model that does not capture the layered nature of the vortex system.

In our simulations, the temperature dependence of the total defect density indicates a change in the nucleation behavior upon cooling at a characteristic temperature, $T_{\rm freez}$. The dependence of $T_{\rm freez}/T_{\rm m}$ on vortex density and sample size suggests that the freezing temperature is influenced by both, intrinsic and confinement effects. In contrast, the systematic increase of the healing length (in units of $a$) with the nanocrystal size at fixed density, in agreement with experimental observations, indicates that confinement effects dominate the structural properties of the frozen nanocrystal. This behavior is consistent with a more prominent role of edge-induced bending as the surface-to-volume vortex ratio decreases. Therefore, elasticity and system size play distinct roles in the freezing dynamics of vortex nanocrystals formed upon cooling.

Our results are largely independent of the microscopic specifics of vortex matter, and should therefore capture general features of crystallization in confined interacting particle systems, including colloids in circular traps and Wigner molecules in quantum dots.

\section{Acknowledgments}
We acknowledge M. Schwerter for stimulating
discussions. Work partially supported by the National Council of Scientific and Technical Research
of Argentina (CONICET) through grants No. PIP 2021-1848, No. PIP 112-202001-01294-CO, and  PIET-R IA-CoNSoFi, by the Universidad Nacional de Cuyo
Research Grant No. 06/80020240100305UN, and by Universidad Nacional de San Luis Research Grant PROICO 03-2220 .

\bibliography{biblio}
\end{document}